\documentstyle[aps,eqsecnum]{revtex}

\begin{document}
\title{Gravitational radiation reaction in compact binary systems: \\
Contribution of the quadrupole-monopole interaction}
\author{L\'{a}szl\'{o} \'{A}. Gergely and Zolt\'{a}n Keresztes}
\address{Astronomical Observatory and Department of Experimental Physics,\\
University of Szeged, D\'{o}m t\'{e}r 9, Szeged, H-6720 Hungary}
\maketitle

\begin{abstract}
The radiation reaction in compact spinning binaries on eccentric orbits due
to the quadrupole-monopole interaction is studied. This contribution is of
second post-Newtonian order. As result of the precession of spins the
magnitude $L$ of the orbital angular momentum is not conserved. Therefore a
proper characterization of the perturbed radial motion is provided by the
energy $E$ and angular average $\bar{L}$. As powerful computing tools, the
generalized true and eccentric anomaly parametrizations are introduced. Then
the secular losses in energy and\ magnitude of orbital angular momentum
together with the secular evolution of the relative orientations of the
orbital angular momentum and spins are found for {\it eccentric} orbits by
use of the residue theorem. The circular orbit limit of the energy loss
agrees with Poisson's earlier result.
\end{abstract}


\section{Introduction}

Compact binary systems composed of neutron stars and/or black holes are
among the best candidates of sources emitting gravitational radiation in the
frequency range of the Earth-based interferometric detectors such as the
Laser Interferometric Gravitational Wave Observatory (LIGO) \cite{LIGO},
VIRGO \cite{VIRGO}, GEO \cite{GEO}, and TAMA \cite{TAMA}, and also for the
envisaged Laser Interferometer Space Antenna (LISA)\cite{LISA,LISA1}. Before
the final coalescence of the compact binary due to a cumulative loss in
energy and angular momentum there is a regime where a post-Newtonian
description of the motion and of the gravitational radiation is well suited,
provided it is taken with sufficient accuracy.

As up to the second post-Newtonian (PN) order the energy and total angular
momentum of the system are conserved, and the leading contribution of
gravitational radiation at 2.5 PN is generally agreed to decrease the
eccentricity of orbits  faster than their radii \cite{Peters}, it is
customary to compute radiation templates and the radiation reaction with the
assumption of circular orbits. Another argument for such a simplifying
assumption is the reduction of the number of fitting parameters in the
matched filtering. However as described in Refs. \cite{QuSha} and \cite{HiBe}%
, there are various astrophysical situations in galactic nuclei where the
eccentricity of the orbits is still significant. The error one makes in
searching for waves emitted by binaries on circular orbits when the waves
come from binaries on eccentric orbits was estimated to be substantial in 
\cite{PoMa}. Therefore it is desirable to have a generic treatment, also
valid for eccentric orbits. Such a treatment, valid up to the second
post-Newtonian order was provided by Gopakumar and Iyer \cite{GI}.

But the situation is even more complicated. At 1.5 PN order other features
of the binary appear. These are spin-orbit type contributions, which further
increase the number of parameters to be fitted. At 2 PN both spin-spin and
quadrupole-monopole effects appear. The latter are the consequences of
considering each component of the binary as a monopole in the quadrupolar
field of the other. The nonradiative dynamics of spinning binaries with
emphasis on both the spins and the quadrupole-monopole interaction was
investigated long ago in a series of papers by Barker and O'Connell (see,
for example, \cite{BOC}). Radiative aspects concerning spins were studied in
Refs. \cite{ACST}-\cite{spinspin2}. However the quadrupole-monopole type
radiative contribution has received less attention until now, excepting the
circular orbit limit, discussed in detail by Poisson \cite{Poisson}. In this
paper we extend this treatment to eccentric orbits.

For this purpose in Sect. II we derive two parametrizations of the orbit,
which generalize the Keplerian true and eccentric anomaly parametrizations
for the case of the quadrupole-monopole perturbation. These parametrizations
have the advantageous property that the required integral expressions can be
easily computed by use of the residue theorem and, further, in the majority
of cases the only pole is at the origin \cite{param}. Our treatment follows
closely and relies on the results of Refs. \cite{spinspin1} and \cite
{spinspin2}. Here we also find the peculiar feature encountered previously
solely in the discussion of the spin-spin contributions, that the magnitude
of the orbital angular momentum is not a conserved quantity. We bypass this
difficulty by introducing its angular average $\bar{L}$ and characterizing
the perturbed orbit by $E$ and $\bar{L}$. It is equally possible to
characterize the motion by $E$ and the time average $\langle L\rangle $;
however the resulting expressions would be more cumbersome. For
completeness, the relation between $\langle L\rangle $ and $\bar{L}$ is
given. At the end of the section we find the same functional expression for
the period of the radial motion as in the unperturbed case, but with the
energy characterizing the perturbed motion. This is a consequence of the
specific functional form of the perturbing term in the radial equation \cite
{param}.

Having developed the required toolchest, the computation of the
quadrupole-monopole contributions to the losses of energy and magnitude of
orbital angular momentum together with the evolution of the angle variables,
which characterize the relative orientation of the spin vectors and orbital
angular momentum vector are straightforward. These are given in Sect. III as
the main results of the paper. We will show in the concluding remarks that
in the circular limit the energy loss reduces to the expression given
previously by Poisson.

The velocity of light $c$ and the gravitational constant $G$ are kept in all
expressions.

\section{The radial motion}

The Keplerian motion of the binary perturbed by the monopole-quadrupole
interaction between its components is properly characterized by the
Lagrangian \cite{Poisson}, \cite{Goldstein}:

\begin{eqnarray}
{\cal L} &=&{\cal L}_{N}+{\cal L}_{QM}\ ,  \nonumber \\
{\cal L}_{N} &=&\frac{\mu {\bf v}^{2}}{2}+\frac{Gm\mu }{r}{\ ,}  \nonumber \\
{\cal L}_{QM} &=&\frac{G\mu m^{3}}{2r^{5}}\sum_{i=1}^{2}p_{i}\left[ 3\left( 
{\bf \hat{S}}_{{\bf i}}\cdot {\bf r}\right) ^{2}-r^{2}\right] \ ,
\label{LagQM}
\end{eqnarray}
where ${\cal L}_{N}$ represents the Newtonian part of the Lagrangian (with $m
$ the total mass and $\mu $ the reduced mass). The summation is taken over
the two bodies in the quadrupole-monopole interaction term ${\cal L}_{QM}$,
which depends both on the direction ${\bf \hat{S}}_{{\bf i}}$ of the spin of
the $i^{th}$ body ${\bf S}_{{\bf i}}$ and on $p_{i}$, defined as 
\begin{equation}
p_{i}=\frac{Q_{i}}{m_{i}m^{2}}\ .  \label{pQ}
\end{equation}
Here $Q_{i}$ is the quadrupole-moment scalar \cite{Poisson} of the $\ i^{%
\text{th}}$ axially symmetric binary component with symmetry axis ${\bf 
\hat{S}}_{{\bf i}}$. In the Newtonian limit computation gives 
\begin{equation}
Q_{i}=\Theta _{i}^{\prime }-\Theta _{i}=-\frac{S_{i}}{\Omega _{i}}\left( 
\frac{\Theta _{i}}{\Theta _{i}^{\prime }}-1\right) \   \label{Qtheta}
\end{equation}
for each axisymmetric body characterized by the principal moments of inertia 
$\left( \Theta _{i},\ \Theta _{i},\ \Theta _{i}^{\prime }\right) $ and
angular velocity $\Omega _{i}=S_{i}/\Theta _{i}^{\prime }$.

Other contributions, like PN, 2PN \cite{GI}, spin-orbit (SO) \cite{GPV3} and
spin-spin (SS) \cite{spinspin1}, \cite{spinspin2} appearing at $1.5$ and
second post-Newtonian orders, respectively will add to the dynamics. However
these other post-Newtonian terms will not interfere with the
quadrupole-monopole interaction terms. Quadrupole-monopole terms are singled
out from among the rest of the second post-Newtonian order terms by the
distinguishing parameters $p_{i}$ in the same way spin-spin interaction
terms can be recognized as being quadratic in the spin magnitudes. Therefore
up to the second post-Newtonian order each contribution can be computed
independently, including the quadrupole-monopole one. In fact the first
interference terms appear in the$\ 2.5$ order, as the first PN correction of
the SO-terms \cite{Owen}.

The acceleration resulting from Eq. (\ref{LagQM}) is 
\begin{eqnarray}
{\bf a} &=&{\bf a}_{N}+{\bf a}_{QM}\ ,  \nonumber \\
{\bf a}_{N} &=&-\frac{Gm{\bf r}}{r^{3}}\ ,  \nonumber \\
{\bf a}_{QM} &=&-\frac{3Gm^{3}}{2r^{7}}\sum_{i=1}^{2}p_{i}\left[ 5\left( 
{\bf \hat{S}}_{{\bf i}}\cdot {\bf r}\right) ^{2}-r^{2}\right] {\bf r} 
\nonumber \\
&&{\bf +}\frac{3Gm^{3}}{r^{5}}\sum_{i=1}^{2}p_{i}\left( {\bf \hat{S}}_{{\bf i%
}}\cdot {\bf r}\right) {\bf \hat{S}}_{{\bf i}}\ .  \label{accQM}
\end{eqnarray}
By performing the required changes in notations the acceleration ${\bf a}%
_{QM}$ coincides with Eqs. (55)-(56) of Ref. \cite{BOC}.

The Lagrangian governing the quadrupole-monopole interaction is not time
dependent; therefore the total energy $E$ of the system is conserved. We
further remark that, as it happens for spin-spin interaction, here the
Lagrangian ${\cal L}_{QM}$ has no dependence on velocities either. Therefore 
${\bf p}=\mu {\bf \dot{r}}$ and two conclusions emerge:

(a.) The energy is given by 
\begin{equation}
E=E_{N}-{\cal L}_{QM}\ ,
\end{equation}

(b.) There is no quadrupole-monopole contribution to the orbital angular
momentum 
\begin{equation}
{\bf L=r\times p=L}_{N}\ .
\end{equation}
The functional form of the Newtonian energy and orbital angular momentum
vector \cite{spinspin1} allows one to relate them to $v^{2}$ and $\dot{r}^{2}
$: 
\begin{eqnarray}
\ v^{2} &=&\frac{2E_{N}}{\mu }+\frac{2Gm}{r}\ ,  \label{v2} \\
\dot{r}^{2} &=&\frac{2E_{N}}{\mu }+\frac{2Gm}{r}-\frac{L_{N}^{2}}{\mu
^{2}r^{2}}\ .  \label{rdot2}
\end{eqnarray}
\qquad 

The cross product of Eq. (\ref{accQM}) with $\mu {\bf r}$ gives the
evolution of the orbital angular momentum:

\begin{equation}
{\bf \dot{L}=}\frac{3G\mu m^{3}}{r^{5}}\sum_{i=1}^{2}p_{i}{\bf \left( \hat{S}%
_{i}\cdot r\right) }\left( {\bf r\times \hat{S}_{i}}\right) {\bf \ .}
\label{Lvecdot}
\end{equation}
Due to the quadrupole-monopole interaction the spins undergo a precessional
motion, computed a long time ago by Barker and O'Connell [Eqs. (39) and (43)
of \cite{BOC} referring to bodies with arbitrary, but constant mass, spin
and quadrupole moment]: 
\begin{equation}
{\bf \dot{S}}_{{\bf i}}=-\frac{3G\mu m^{3}}{r^{5}}p_{i}{\bf \left( \hat{S}%
_{i}\cdot r\right) }\left( {\bf r}\times {\bf \hat{S}}_{{\bf i}}\right) \ .
\label{Sprec}
\end{equation}
It is then straightforward to check the conservation of the total angular
momentum 
\begin{equation}
{\bf J=L+S}_{{\bf 1}}+{\bf S}_{{\bf 2}}\ .  \label{J}
\end{equation}
\qquad \qquad 

Multiplying Eq. (\ref{Lvecdot}) by the direction of the orbital angular
momentum ${\bf \hat{L}}$ the magnitude of the angular momentum is found to
evolve. The nonconservation of the magnitude of the orbital angular momentum
is a feature already encountered when studying spin-spin dynamics. By
performing the transformations enlisted in Ref \cite{spinspin1} the detailed
form of the evolution of $L$ emerges

\begin{equation}
\dot{L}=-\frac{3G\mu m^{3}}{2r^{3}}\sum_{i=1}^{2}p_{i}\sin ^{2}\kappa
_{i}\sin 2(\psi -\psi _{i})\ .  \label{LdotAng}
\end{equation}
Here and hereafter we employ the following angles. $\kappa _{i}=\cos ^{-1}(%
{\bf \hat{S}_{i}\cdot \hat{L})}$ and $\gamma =\cos ^{-1}({\bf \hat{S}%
_{1}\cdot \hat{S}_{2})}$ characterize the relative orientation of the
angular momenta. The angles $\psi $ and$\ \psi _{i}$ are subtended by the
intersection line of the planes perpendicular to ${\bf J}$ and ${\bf L}\ $%
with the position ${\bf r}$ and the projections of the spins in the plane of
the orbit, respectively [see Fig. 1 in Ref. \cite{GPV3}. We also introduce $%
\delta _{i}=2\left( \psi _{0}-\psi _{i}\right) $]$.$ This description allows
for computing the value of $L$ in terms of the Newtonian true anomaly
parameter $\chi =\psi -\psi _{0}$ and a properly defined angular average $%
\bar{L}$ of $L$ in a similar way as was done in the spin-spin case: 
\begin{eqnarray}
L(\chi ) &=&\bar{L}+\delta L  \nonumber \\
\delta L &=&\frac{G\mu ^{3}m^{3}}{4\bar{L}^{3}}\!\!\sum_{i=1}^{2}p_{i}\sin
^{2}\!\kappa _{i}\{2\bar{A}\cos \!\left( \chi \!+\!\delta _{i}\right)  
\nonumber \\
&&\!+\!\left( 3G\mu m\!+\!2\bar{A}\cos \!\chi \right) \cos (\!2\!\chi
\!+\!\delta _{i})\}\ .  \label{Lchi}
\end{eqnarray}
where 
\begin{equation}
\bar{A}=\left( G^{2}m^{2}\mu ^{2}+\frac{2E\bar{L}^{2}}{\mu }\right) ^{1/2}
\label{Abar}
\end{equation}
is the magnitude of the Laplace-Runge-Lenz vector for a Keplerian motion
characterized by $E$ and $\bar{L}$.

By following the method of Ref. \cite{spinspin1} the quadrupole-monopole
interaction part $E_{QM}$ of the energy can also be given in terms of $\chi $
and $r$ : 
\begin{equation}
E_{QM}(r,\chi )=\frac{G\mu m^{3}}{2r^{3}}\!\!\sum_{i=1}^{2}p_{i}\left[
1-\!3\!\sin ^{2}\!\kappa _{i}\cos ^{2}\!\left( \!\chi \!+\frac{\delta _{i}}{2%
}\!\right) \right] \ ,  \label{Echir}
\end{equation}
Then the expressions (\ref{v2}) and (\ref{rdot2}) take the form: 
\begin{eqnarray}
\ v^{2} &=&\frac{2[E-E_{QM}(r,\chi )]}{\mu }+\frac{2Gm}{r}\ ,  \label{v2Ang}
\\
\dot{r}^{2} &=&\frac{2[E-E_{QM}(r,\chi )]}{\mu }+\frac{2Gm}{r}-\frac{L(\chi
)^{2}}{\mu ^{2}r^{2}}\ ,  \label{rdot2Ang}
\end{eqnarray}
with $L(\chi )$ and $E_{QM}(r,\chi )$ given by Eqs. (\ref{Lchi}) and (\ref
{Echir}). As the parameter $\chi $ appears only in second post-Newtonian
terms, it can be replaced by any generalized true anomaly parameter having
the proper Newtonian limit. In what follows we will seek for such a suitable
parametrization $r=r\left( \chi \right) $ of the radial component of the
perturbed motion.

For this purpose we compute the turning points of the orbit, defined as
roots of the radial equation (\ref{rdot2Ang}) with $\dot{r}^{2}=0$ for $\chi
=0,\pi $. We find:

\begin{eqnarray}
r_{{}_{{}_{min}^{max}}} &=&\frac{Gm\mu \pm \bar{A}}{-2E}+\frac{G\mu ^{2}m^{3}%
}{4\bar{A}\bar{L}^{2}}\sum_{i=1}^{2}p_{i}\rho _{\mp }^{i}\ ,  \nonumber \\
\rho _{\mp }^{i} &=&\alpha _{0}^{i}(\bar{A}\mp Gm\mu )+\beta _{0}^{i}(4%
\bar{A}\mp 3Gm\mu )\ ,  \nonumber \\
\alpha _{0}^{i} &=&2\left[ 1-\!3\!\sin ^{2}\!\kappa _{i}\cos ^{2}\left( \psi
_{0}\!-\!\psi _{i}\right) \!\right] \ ,  \nonumber \\
\beta _{0}^{i} &=&\sin ^{2}\kappa _{i}\cos 2(\psi _{0}-\psi _{i})\ .
\label{turningpoints}
\end{eqnarray}
Then, following Ref. \cite{spinspin1} we derive the generalized true anomaly
parametrization

\begin{equation}
r=\frac{\bar{L}^{2}}{\mu (Gm\mu +\bar{A}\cos \chi )}+\frac{G\mu ^{2}m^{3}}{4%
\bar{A}\bar{L}^{2}(Gm\mu +\bar{A}\cos \chi )^{2}}\sum_{i=1}^{2}p_{i}\Lambda
^{i}\ ,  \label{param}
\end{equation}
with 
\begin{eqnarray}
\Lambda ^{i} &=&\bar{A}[\bar{A}^{2}(\alpha _{0}^{i}\!+\!4\beta
_{0}^{i})\!+\!(Gm\mu )^{2}(3\alpha _{0}^{i}\!+\!10\beta _{0}^{i})]  \nonumber
\\
&+&\!Gm\mu \lbrack \bar{A}^{2}(3\alpha _{0}^{i}\!+\!11\beta
_{0}^{i})\!+\!(Gm\mu )^{2}(\alpha _{0}^{i}\!+\!3\beta _{0}^{i})]\cos \!\chi
\ .
\end{eqnarray}
In a similar way we derive the generalized eccentric anomaly parametrization 
\begin{equation}
r=\frac{Gm\mu -\bar{A}\cos \xi }{-2E}+\frac{G\mu ^{2}m^{3}}{4\bar{A}\bar{L}%
^{2}}\sum_{i=1}^{2}p_{i}\Xi ^{i}\ ,  \label{paramxi}
\end{equation}
with 
\begin{equation}
\Xi ^{i}=\bar{A}(\alpha _{0}^{i}+4\beta _{0}^{i})+Gm\mu (\alpha
_{0}^{i}+3\beta _{0}^{i})\cos \!\xi \ .
\end{equation}
Note that $\rho _{\mp }^{i}$, $\Lambda ^{i}$, and $\Xi ^{i}$ have the same
structure as in the spin-spin case; however, the angular expressions $\alpha
_{0}^{i}\!$ and $\beta _{0}^{i}$ are different from $\alpha _{0}\!$ and $%
\beta _{0}$ introduced in Ref. \cite{spinspin1}.

In the perturbative terms we will need the Keplerian relations 
\begin{equation}
\dot{r}=\frac{\bar{A}}{\bar{L}}\sin \chi \ ,\qquad \dot{\psi}=\frac{\bar{L}}{%
\mu r^{2}}\ .  \label{Kep}
\end{equation}

The two parametrizations are suitable for averaging radial expressions by
use of the residue theorem, as described in detail in \cite{param}. The time
average of $L(\chi )$ is 
\begin{equation}
\langle L\rangle =\bar{L}+\frac{Gm^{3}\mu ^{2}F_{1}}{4\bar{A}^{2}\bar{L}%
^{3}F_{2}}\sum_{i=1}^{2}p_{i}\sin ^{2}\kappa _{i}\cos \delta _{i}
\label{LaveLbar}
\end{equation}
with the coefficients $F_{1,2}$ given by 
\begin{eqnarray}
F_{1} &=&2\bar{L}(-2\mu E)^{1/2}[\bar{A}^{6}-15G^{2}m^{2}\mu ^{2}\bar{A}%
^{4}+32G^{4}m^{4}\mu ^{4}\bar{A}^{2}-16G^{6}m^{6}\mu ^{6}]  \nonumber \\
&&+Gm\mu ^{2}[-11\bar{A}^{6}+58G^{2}m^{2}\mu ^{2}\bar{A}^{4}-80G^{4}m^{4}\mu
^{4}\bar{A}^{2}+32G^{6}m^{6}\mu ^{6}]  \nonumber \\
F_{2} &=&4Gm\bar{L}(-2\mu E)^{1/2}[\bar{A}^{2}-2G^{2}m^{2}\mu ^{2}]+\bar{A}%
^{4}-8G^{2}m^{2}\mu ^{2}\bar{A}^{2}+8G^{4}m^{4}\mu ^{4}\ .
\end{eqnarray}
We remark that the coefficients $F_{1,2}$ are the same as the corresponding
coefficients in the spin-spin case. As $L\left( \chi \right) $ given in
terms of $\langle L\rangle $ has a more cumbersome expression than Eq. (\ref
{Lchi}), we will give all forthcoming expressions in terms of $\bar{L}$.

For the period once again we find a Keplerian expression, but with $E$
characterizing the perturbed dynamics: 
\begin{equation}
T=2\pi Gm\left( \frac{\mu }{-2E}\right) ^{3/2}\ .  \label{period}
\end{equation}

\section{Leading order quadrupole-monopole contribution to the evolutions of
the dynamical variables under radiation reaction}

\subsection{Energy loss}

To leading order the instantaneous loss of energy under radiation reaction
is given by Einstein's quadrupole formula 
\begin{equation}
\frac{dE}{dt}=-\frac{G}{5c^{5}}I^{(3)jl}I^{(3)jl}\ ,  \label{quadrup}
\end{equation}
where $\epsilon ^{ijk}$ is the completely antisymmetric Levi-Civita symbol,
the numbers in parantheses denote a corresponding order time derivative and
the {\it system}'s symmetric trace-free (STF) mass quadrupole moment $%
I_{N}^{\ jl}$ to leading order is given by 
\begin{equation}
I_{N}^{\ jl}=\mu \left( x^{j}x^{l}\right) ^{STF}\ .  \label{massquad}
\end{equation}
The Newtonian and quadrupole-monopole contribution to the instantaneous loss
under radiation reaction in the energy is : 
\begin{equation}
\frac{dE}{dt}=-\frac{G}{5c^{5}}I_{N}^{(3)jl}({\bf a}_{N})I_{N}^{(3)kl}({\bf a%
}_{N})-\frac{2G}{5c^{5}}I_{N}^{(3)jl}({\bf a}_{QM})I_{N}^{(3)kl}({\bf a}%
_{N})\ .  \label{ElossK}
\end{equation}
The arguments of time derivatives of the momenta contain the contribution to
the acceleration to be inserted in the respective terms.\ By inserting in
the Newtonian terms the expressions (\ref{v2Ang}) and (\ref{rdot2Ang}) for $%
v^{2}$ and $\dot{r}^{2}$, respectively, and $L(\chi )$ and $E_{QM}(r,\chi )$
given by Eqs. (\ref{Lchi}) and (\ref{Echir}) and the Newtonian expressions
in the quadrupole-monopople terms, we obtain:

\begin{eqnarray}
\frac{dE}{dt} &=&\left( \frac{dE}{dt}\right) _{N}+\left( \frac{dE}{dt}%
\right) _{QM}\ ,  \label{ElossInst} \\
\left( \frac{dE}{dt}\right) _{N} &=&-\frac{8G^{3}m^{2}}{15c^{5}r^{6}}(2\mu
E\!r^{2}\!+\!2Gm\mu ^{2}r\!+\!11\bar{L}^{2})\ ,  \label{ElossInstN} \\
\left( \frac{dE}{dt}\right) _{QM} &=&\frac{2G^{3}m^{4}}{15c^{5}\bar{L}%
^{2}r^{8}}\sum_{i=1}^{2}p_{i}\left\{ \sum_{n=1}^{3}a_{n}\cos \left( n\chi
+\delta _{i}\right) \sin ^{2}\kappa _{i}+a_{4}(2-3\sin ^{2}\kappa
_{i})\right\} \ .  \label{ElossinstQM}
\end{eqnarray}
The coefficients $a_{k}$ are given by 
\begin{eqnarray}
a_{1} &=&3\mu \bar{A}r(-22Gm\mu ^{2}r\!+17\bar{L}^{2})\ ,  \nonumber \\
a_{2} &=&6(-11G^{2}m^{2}\mu ^{4}r^{2}+6E\bar{L}^{2}\mu r^{2}+5Gm\mu ^{2}\bar{%
L}^{2}r\!-51\bar{L}^{4})\ ,  \nonumber \\
a_{3} &=&-\mu \bar{A}r(22Gm\mu ^{2}r\!+51\bar{L}^{2})\ ,  \nonumber \\
a_{4} &=&2\bar{L}^{2}(-6E\mu r^{2}-5Gm\mu ^{2}r\!+39\bar{L}^{2})\ .
\end{eqnarray}
By use of the true anomaly parametrization $r(\chi )$, Eq. (\ref{param}) we
find the energy loss in terms of $\chi $ alone. Then we pass to the complex
parameter $z=\exp (i\chi )$ and we compute the averaged energy loss by use
of the residue theorem, the only pole being at the origin: 
\begin{eqnarray}
\left\langle \frac{dE}{dt}\right\rangle  &=&\left\langle \frac{dE}{dt}%
\right\rangle _{N}+\left\langle \frac{dE}{dt}\right\rangle _{QM}\ ,
\label{AvelossE} \\
\left\langle \frac{dE}{dt}\right\rangle _{N} &=&-\frac{G^{2}m(-2E\mu )^{3/2}%
}{15c^{5}\bar{L}^{7}}(148E^{2}\bar{L}^{4}+732G^{2}m^{2}\mu ^{3}E\bar{L}%
^{2}+425G^{4}m^{4}\mu ^{6})\ ,  \label{AvelossEN} \\
\left\langle \frac{dE}{dt}\right\rangle _{QM} &=&\frac{G^{2}\mu m^{3}(-2E\mu
)^{3/2}}{30c^{5}\bar{L}^{11}}\sum_{i=1}^{2}p_{i}[C_{1}\sin ^{2}\kappa
_{i}\cos \delta _{i}+C_{2}(2-3\sin ^{2}\kappa _{i})]\ ,  \label{AvelossEQM}
\end{eqnarray}
with the coefficients $C_{k}$ given by: 
\begin{eqnarray}
C_{1} &=&-\mu \bar{A}^{2}(948E^{2}\bar{L}^{4}+8936G^{2}m^{2}\mu ^{3}E\bar{L}%
^{2}+8335G^{4}m^{4}\mu ^{6})\ ,  \nonumber \\
C_{2} &=&708E^{3}\bar{L}^{6}+10020G^{2}m^{2}\mu ^{3}E^{2}\bar{L}%
^{4}+18865G^{4}m^{4}\mu ^{6}E\bar{L}^{2}+8316G^{6}m^{6}\mu ^{9}\ .
\label{C12}
\end{eqnarray}

\subsection{Change in the magnitude of orbital angular momentum}

The instantaneous and averaged losses in $L$ under the radiation reaction
can be found in the same way as for the spin-spin case \cite{spinspin1}: 
\begin{equation}
\frac{dL}{dt}={\bf \hat{L}\cdot }\frac{d{\bf J}}{dt}-{\bf \hat{L}\cdot }%
\frac{d{\bf S_{1}}}{dt}-{\bf \hat{L}\cdot }\frac{d{\bf S_{2}}}{dt}\ .
\end{equation}
\qquad 

The evolution under radiation reaction in the spin ${\bf S_{i}}$ of the $%
i^{th}$ axisymmetric body (or {\it approximately} axisymmetric, with the
deviation from axisymmetry of any post-Newtonian order), following \cite
{ACST} was derived in \cite{GPV2} by computing the integral of the moment of
the reaction force (the sign swapped gradient of the Burke-Thorne potential)
over the volume of each body. The radiation reaction will change the
instantaneous orientation but not the magnitude of the spin vectors. However
in \cite{spinspin2} it was shown that there is no{\it \ secular} spin
evolution under radiation reaction in the $2PN$ order{\it :} 
\begin{equation}
\left\langle \frac{d{\bf S_{i}}}{dt}\right\rangle =0\ .  \label{nospinave}
\end{equation}
Denoting by $\simeq $ equalities modulo Burke-Thorne type terms, e.g.
disregarding all terms which average out due to Eq.(\ref{nospinave}) we\
find for the instantaneous loss in $L$ under the radiation reaction: 
\begin{equation}
\frac{dL}{dt}\simeq {\bf \hat{L}\cdot }\frac{d{\bf J}}{dt}\ .
\end{equation}
The leading order instantaneous loss of ${\bf J}$ under radiation reaction
is: 
\begin{equation}
\frac{d{\bf J}^{i}}{dt}=-\frac{2G}{5c^{5}}\epsilon ^{ijk}I^{(2)jl}I^{(3)kl}\
.  \label{quadrupJ}
\end{equation}
The Newtonian and quadrupole-monopole terms decouple as: 
\begin{equation}
\frac{d{\bf J}^{i}}{dt}=-\frac{2G}{5c^{5}}\epsilon ^{ijk}I_{N}^{(2)jl}({\bf a%
}_{N})I_{N}^{(3)kl}({\bf a}_{N})-\frac{2G}{5c^{5}}\epsilon ^{ijk}\left[
I_{N}^{(2)jl}({\bf a}_{N})I_{N}^{(3)kl}({\bf a}_{QM})+I_{N}^{(2)jl}({\bf a}%
_{QM})I_{N}^{(3)kl}({\bf a}_{N})\right] \ .
\end{equation}
In the QM-terms we can use Cartesian coordinates $(x,y,z)=r(\cos \psi ,\sin
\psi ,0)$ [equivalent to ${\bf \hat{L}=(}0,0,1)$]. Then the spins are
expressed as ${\bf S_{i}=}S_{i}(\sin \kappa _{i}\cos \psi _{i},\sin \kappa
_{i}\sin \psi _{i},\cos \kappa _{i}).$ The computation yields the following
instantaneous losses: 
\begin{eqnarray}
{\bf \hat{L}\cdot }\frac{d{\bf J}}{dt} &=&\left( {\bf \hat{L}\cdot }\frac{d%
{\bf J}}{dt}\right) _{N}+\left( {\bf \hat{L}\cdot }\frac{d{\bf J}}{dt}%
\right) _{QM}\ ,  \label{Llossinst} \\
\left( {\bf \hat{L}\cdot }\frac{d{\bf J}}{dt}\right) _{N} &=&\frac{8G^{2}m%
\bar{L}}{5c^{5}\mu r^{5}}\left( 2\mu Er^{2}-3\bar{L}^{2}\right) \ ,
\label{LlossN} \\
\left( {\bf \hat{L}\cdot }\frac{d{\bf J}}{dt}\right) _{QM} &=&\frac{%
2G^{2}m^{3}}{5c^{5}\mu \bar{L}^{3}r^{7}}\sum_{i=1}^{2}p_{i}\left\{
\sum_{n=1}^{3}b_{n}\cos \left( n\chi +\delta _{i}\right) \sin ^{2}\kappa
_{i}+b_{4}(2-3\sin ^{2}\kappa _{i})\right\} \ ,  \label{LlossQM}
\end{eqnarray}
with the coefficients $b_{k}$ given by 
\begin{eqnarray}
b_{1} &=&3\mu \bar{A}r(2Gm\mu ^{3}Er^{3}-\mu E\bar{L}^{2}r^{2}-9Gm\mu ^{2}%
\bar{L}^{2}r+8\bar{L}^{4})\ ,  \nonumber \\
b_{2} &=&3(2G^{2}m^{2}\mu ^{5}Er^{4}-9G^{2}m^{2}\mu ^{4}\bar{L}%
^{2}r^{2}+22\mu E\bar{L}^{4}r^{2}+10Gm\mu ^{2}\bar{L}^{4}r-19\bar{L}^{6})\ ,
\nonumber \\
b_{3} &=&\mu \bar{A}r(2Gm\mu ^{3}Er^{3}+3\mu E\bar{L}^{2}r^{2}-9Gm\mu ^{2}%
\bar{L}^{2}r-24\bar{L}^{4})\ ,  \nonumber \\
b_{4} &=&\bar{L}^{4}(-18\mu Er^{2}-8Gm\mu ^{2}r+15\bar{L}^{2})\ .
\end{eqnarray}
A similar averaging procedure as in the case of the energy loss, gives here 
\begin{eqnarray}
\left\langle \frac{dL}{dt}\right\rangle  &=&\left\langle \frac{dL}{dt}%
\right\rangle _{N}+\left\langle \frac{dL}{dt}\right\rangle _{QM}\ ,
\label{avelossL} \\
\left\langle \frac{dL}{dt}\right\rangle _{N} &=&-\frac{4G^{2}m(-2\mu E)^{3/2}%
}{5c^{5}\bar{L}^{4}}(14E\bar{L}^{2}+15G^{2}m^{2}\mu ^{3})\ ,
\label{avelossLN} \\
\left\langle \frac{dL}{dt}\right\rangle _{QM} &=&\frac{G^{2}m^{3}\mu (-2\mu
E)^{3/2}}{10c^{5}\bar{L}^{8}}\sum_{i=1}^{2}p_{i}[D_{1}\sin ^{2}\kappa
_{i}\cos \delta _{i}+D_{2}(2-3\sin ^{2}\kappa _{i})]\ ,  \label{avelossLQM}
\end{eqnarray}
where the coefficients $D_{k}$ are given below 
\begin{eqnarray}
D_{1} &=&-6(62E^{2}\bar{L}^{4}+211G^{2}m^{2}\mu ^{3}E\bar{L}%
^{2}+90G^{4}m^{4}\mu ^{6})\ ,  \nonumber \\
D_{2} &=&252E^{2}\bar{L}^{4}+1200G^{2}m^{2}\mu ^{3}E\bar{L}%
^{2}+805G^{4}m^{4}\mu ^{6}\ .
\end{eqnarray}

\subsection{Evolution of angles under radiation reaction}

The equations for the evolution under radiation reaction of the angles $%
\kappa _{i}=\cos ^{-1}({\bf \hat{S}_{i}\cdot \hat{L})},\ (i=1,2)$ and $%
\gamma =\cos ^{-1}({\bf \hat{S}_{1}\cdot \hat{S}_{2})}$ were derived in \cite
{spinspin2}:

\begin{eqnarray}
\frac{d}{dt}\cos \gamma  &\simeq &0\ ,  \label{gammaloss2} \\
\frac{d}{dt}\cos \kappa _{i} &\simeq &\left( \frac{d}{dt}\cos \kappa
_{i}\right) _{SO}+\left( \frac{d}{dt}\cos \kappa _{i}\right) _{SS}+\left( 
\frac{d}{dt}\cos \kappa _{i}\right) _{QM}\ .  \label{kappaloss2}
\end{eqnarray}
The averaged $SO$ terms were computed previously and given in \cite{GPV3}
and the averaged $SS$ terms in \cite{spinspin2}. To the expression derived
in \cite{spinspin2} we have added the required quadrupole-monopole term: 
\begin{equation}
\left( \frac{d}{dt}\cos \kappa _{i}\right) _{QM}\simeq \frac{1}{\bar{L}}(%
{\bf \hat{S}}_{{\bf i}}-{\bf \hat{L}}\cos \kappa _{i})\cdot \left( \frac{d%
{\bf J}}{dt}\right) _{QM}\ ,  \label{kappaloss4}
\end{equation}
This is the new contribution to be computed in this subsection. For this
purpose we rewrite the expression (\ref{kappaloss4}) as a function of the
radial variables $r(\chi )$ and $\chi :$%
\begin{eqnarray}
\frac{1}{\bar{L}}({\bf \hat{S}}_{{\bf i}}-{\bf \hat{L}}\cos \kappa
_{i})\cdot \left( \frac{d{\bf J}}{dt}\right) _{QM} &=&\frac{3G^{2}m^{3}}{%
10c^{5}\mu \bar{L}^{2}r^{7}}\sin \kappa _{i}\sum_{j=1}^{2}p_{j}\sin 2\kappa
_{j}\Bigl\{\sum_{n=1}^{3}u_{n}\cos \left( n\chi +\frac{\delta _{i}+\delta
_{j}}{2}\right)   \nonumber \\
&&+u_{4}\sin \chi \sin \left( \frac{\delta _{i}-\delta _{j}}{2}\right)
+u_{5}\cos \left( \frac{\delta _{i}-\delta _{j}}{2}\right) \Bigr\}\ ,
\end{eqnarray}
\begin{eqnarray}
u_{1} &=&-u_{3}=\mu \bar{A}r(2\mu Er^{2}-3\bar{L}^{2})\ ,  \nonumber \\
u_{2} &=&-2\bar{L}^{2}(Gm\mu ^{2}r+3\bar{L}^{2})\ ,  \nonumber \\
u_{4} &=&2\mu \bar{A}r(2\mu Er^{2}-5\bar{L}^{2})\ ,  \nonumber \\
u_{5} &=&2\bar{L}^{2}(4\mu Er^{2}+Gm\mu ^{2}r-5\bar{L}^{2})\ .
\end{eqnarray}

The residue theorem yields the averages: 
\begin{eqnarray}
\left\langle \frac{d\gamma }{dt}\right\rangle  &=&0  \label{gammaloss} \\
\left\langle \frac{d\kappa _{i}}{dt}\right\rangle  &=&\left\langle \frac{%
d\kappa _{i}}{dt}\right\rangle _{SO}+\left\langle \frac{d\kappa _{i}}{dt}%
\right\rangle _{SS-self}+\left\langle \frac{d\kappa _{i}}{dt}\right\rangle
_{SS}+\left\langle \frac{d\kappa _{i}}{dt}\right\rangle _{QM}
\label{kappaloss} \\
\left\langle \frac{d\kappa _{i}}{dt}\right\rangle _{SO} &&\ given\ by\ Eq.\
(4.4)\ of\ \cite{GPV3}  \label{kappalossSO} \\
\left\langle \frac{d\kappa _{i}}{dt}\right\rangle _{SS-self} &&\ given\ by\
Eq.\ (4.7)\ of\ \cite{spinspin2}  \label{kappalossself} \\
\left\langle \frac{d\kappa _{i}}{dt}\right\rangle _{S_{1}S_{2}} &&\ given\
by\ Eq.\ (4.8)\ of\ \cite{spinspin2}\ ,  \label{kappalossSS} \\
\left\langle \frac{d\kappa _{i}}{dt}\right\rangle _{QM} &=&\frac{%
3G^{2}m^{3}\mu (-2\mu E)^{3/2}}{10c^{5}\bar{L}^{9}}\sum_{j=1}^{2}p_{j}\sin
2\kappa _{j}\left[ V_{1}\cos \left( \frac{\delta _{i}+\delta _{j}}{2}\right)
+V_{2}\cos \left( \frac{\delta _{i}-\delta _{j}}{2}\right) \right] \ ,
\label{kappalossQM}
\end{eqnarray}
\ where the coefficients $V_{1-3}$ are: 
\begin{eqnarray}
V_{1} &=&40E^{2}\bar{L}^{4}+90G^{2}m^{2}\mu ^{3}E\bar{L}^{2}+35G^{4}m^{4}\mu
^{6}  \nonumber \\
V_{2} &=&48E^{2}\bar{L}^{4}+140G^{2}m^{2}\mu ^{3}E\bar{L}^{2}+70G^{4}m^{4}%
\mu ^{6}\ .
\end{eqnarray}
For consistency in notation the replacements $L\rightarrow \bar{L}$ and $%
A_{0}\rightarrow \bar{A}$ should be carried out in Eq. (4.4)$\ $of \cite
{GPV3} .

Equations (\ref{gammaloss})-(\ref{kappalossQM}) give the{\it \ complete
evolution under radiation reaction of the angles characterizing the relative
orientation of the spin and orbital angular-momentum vectors up to second
post-Newtonian order. }

\section{Concluding remarks}

As the main result we have derived the second post-Newtonian order
quadrupole-monopole contribution to the radiation reaction for the energy,
the magnitude of the orbital angular momentum, and the angular variables
characterizing the relative orientation of angular momenta of a coalescing
binary system. This was possible by introducing a suitable angular average $%
\bar{L}$ together with the generalized true and eccentric anomaly
parametrizations for the quadrupole-monopole perturbation of the Keplerian
motion.

The secular losses of $E$, Eqs. (\ref{AvelossE})-(\ref{AvelossEQM}) and of $L
$, Eqs. (\ref{avelossL})-(\ref{avelossLQM}), complementing the corresponding
spin-orbit terms of Ref. \cite{GPV3} (with the replacements $L\rightarrow 
\bar{L}$ and $A_{0}\rightarrow \bar{A}$), the spin-spin terms of Ref. \cite
{spinspin1}, and the post-Newtonian terms of Ref. \cite{GI} give the total
radiation reaction up to second post-Newtonian order.

{\it All} contributions to the secular evolutions of the angles $\kappa _{i}$
and $\gamma $ were listed in the Sec III.C.\ They are of either spin or
quadrupolar origin.

Finally, we compare our result for the energy loss with the one given by
Poisson in Ref. \cite{Poisson} for circular orbits. For this we need to take
the circular orbit limit of Eqs. (\ref{AvelossE})-(\ref{C12}). By the
circular orbit limit in Ref. \cite{Poisson} is meant a perturbation of a
circular Keplerian orbit. This, however, cannot be immediately imposed in
our formalism, relying on heavy use of conserved quantities of the perturbed
motion. It is certainly impossible to impose the circularity conditions for
the unperturbed orbit as $E_{N}=E-E_{QM}=-Gm\mu /2r_{0}$ and $%
L_{N}^{2}=L^{2}\left( \chi \right) =(\bar{L}+\delta L)^{2}=Gm\mu ^{2}r_{0}$ (%
$r_{0}$ being the radius of the unperturbed orbit) due to the fact that $%
E_{QM}$ and $\delta L$ are not constants, while $E\ ,\bar{L}$ and $r_{0}$
are. This is related to the remark of \cite{Poisson} that the orbits cannot
be circular in the strict sense. However, we can impose the circularity
conditions in an average sense. As the angular average of $\delta L$
vanishes, we obtain: 
\begin{eqnarray}
\bar{E}_{N} &=&E-\bar{E}_{QM}=-\frac{Gm\mu }{2r_{0}}\ ,  \nonumber \\
\bar{L}_{N}^{2} &=&\bar{L}^{2}=Gm\mu ^{2}r_{0}\ .  \label{cr1}
\end{eqnarray}
These values inserted in the angular average of $E_{QM}\,$\ (which is
already of second PN order) give 
\begin{equation}
\bar{E}_{QM}=\frac{Gm^{3}\mu }{4r_{0}^{3}}\sum_{i=1}^{2}p_{i}\left( 3\cos
^{2}\kappa _{i}-1\right) \ .  \label{cr2}
\end{equation}
Expressing $E$ and $\bar{L}$ from Eqs. (\ref{cr1}) and (\ref{cr2}) and
inserting them into the expressions (\ref{AvelossE})-(\ref{C12}) for the
energy loss we find the radiative energy loss for the above defined circular
orbits: 
\begin{equation}
\left\langle \frac{dE}{dt}\right\rangle =-\frac{32G^{4}m^{3}\mu ^{2}}{%
5c^{5}r_{0}^{5}}\left[ 1-6\frac{m^{2}}{r_{0}^{2}}\sum_{i=1}^{2}p_{i}\left(
3\cos ^{2}\kappa _{i}-1\right) \right] \ .  \label{cr3}
\end{equation}
Employing the relation between the radius of the unperturbed circular orbit $%
r_{0}$ and the average velocity of the perturbed orbit $\langle v\rangle $,
deducible from the expressions given in Ref. \cite{Poisson}: 
\begin{equation}
r_{0}=\frac{Gm}{\langle v\rangle ^{2}}\left[ 1-\frac{\langle v\rangle ^{4}}{%
G^{2}}\sum_{i=1}^{2}p_{i}\left( 3\cos ^{2}\kappa _{i}-1\right) \right] \ ,
\label{cr4}
\end{equation}
we find that the averaged energy loss for circular orbits agrees with Eq.
(22) of Ref. \cite{Poisson}.

\section{Acknowledgments}

This work was supported by the Hungarian Higher Education and Research
Foundation (AMFK). The algebraic package REDUCE was employed in some of the
computations.

\end{document}